\documentclass[aps,prl,twocolumn,showpacs]{revtex4}
\usepackage{amsmath}
\usepackage{graphicx}
\usepackage{units}  
\usepackage{xspace}
 
\usepackage{hyperref}
\usepackage{amssymb}
 \usepackage{amsmath}
\usepackage{graphicx}
\usepackage{epsfig}
\newcommand{\beq}{\begin{equation}}
\newcommand{\eeq}{\end{equation}}
\newcommand{\bea}{\begin{eqnarray}}
\newcommand{\eea}{\end{eqnarray}}

\begin{document}

 \bibliographystyle{apsrev}
 
\title{Topological crystalline insulator phase in graphene multilayers }

\author{M. Kindermann}

\affiliation{
 School of Physics, Georgia Institute of Technology, Atlanta, Georgia 30332, USA}

\begin{abstract}
 While the experimental progress on three dimensional topological insulators  is rapid, the development of their two dimensional counterparts has been comparatively slow, despite their technological promise. The main reason is materials challenges of the to date only realizations of two-dimensional topological insulators, in semiconductor quantum wells. Here  we identify a two dimensional topological insulator in a material which does not face similar challenges and which is by now most widely available and well-charaterized: graphene. For certain commensurate interlayer twists graphene multilayers are insulators with sizable bandgaps. We show that they are moreover in a topological phase protected by crystal symmetry. As its fundamental signature, this topological state supports one-dimensional  boundary modes. They form low-dissipation quantum wires that can be defined purely electrostatically. 
 \end{abstract}

\pacs{73.22.Pr,73.21.Ac,73.43.-f,72.80.Vp}
\maketitle

There are physical systems with properties that depend only on   global topology, but not on local details. Because of their inherent stability against perturbations systems in such topological phases have attracted much interest, both from a fundamental and a technological point of view. The first   topological phase realized in a condensed matter system was the quantum Hall state, which occurs in  two-dimensional electron gases subject to a strong magnetic field. Rather recently a new class of such systems has been discovered: topological insulators (TI) \cite{moore:nat10,hasan:rmp10,qi:rmp11}. These materials  typically have strong spin-orbit interactions and  their topological character invokes time-reversal symmetry. Later, also topological phases that are protected by crystal symmetries  \cite{fu:prl11,slager:nap13}, such as mirror symmetry \cite{hsieh:nac12,tanaka:nap12,dziawa:nam12,Xu:12,yao:12,xu:nac12,zhang:13},  have been predicted and observed, so-called  topological crystalline insulators (TCI).
 
The experimental activity in the field so far has focused on three-dimensional systems, although TIs in two dimensions enjoy a number of advantages. For example, bulk charge carriers that pose a major experimental challenge for their three-dimensional counterparts can be eliminated easily by gating. Nevertheless, experimental progress on two-dimensional TIs to date is hampered by  materials challenges. The only two-dimensional TI realized so far is the quantum spin Hall (QSH) state \cite{qi:rmp11} induced by spin-orbit interaction and characterized by pairs of time-reversed boundary modes, so-called helical edge states.  This state was predicted to occur in graphene \cite{kane:prl05,kane:prl05b},   HgTe and InAs/GaSb quantum wells \cite{bernevig:sci06,liu:prl08},  and several other two-dimensional materials   \cite{murakami:prl06,xiao:nac11,ghaemi:prb12}.   It  was observed in milestone experiments on HgTe \cite{Kšnig:sci07} and later in InAs/GaSb \cite{knez:prl11}.   Subsequent research activity, however, has been  limited due to experimental challenges specific to these material systems. Graphene does not pose those challenges and it is a by now  most widely available and investigated material. From this point of view it appears to be the ideal system to implement two-dimensional TI states. However, unfortunately the spin-orbit coupling  in graphene is too weak  \cite{yao:prb07,boettger:prb07} to observe the QSH phase as  originally predicted \cite{kane:prl05} in present day experiments. 
 This has motivated much subsequent research on how to drive graphene into a stable topological phase by  many-body interactions    \cite{raghu:prl08,weeks:prb10,pereg:prb12}, radiation \cite{lindner:nap11}, or enhancing the spin-orbit interaction 
 \cite{varykhalov:prl08,neto:prl09,abdelouahed:prb10,weeks:prx11,qiao:prl12,hu:prl12,qiao:prb13}. 
 
 In this Letter we propose an alternate route to inducing  topological phases in graphene: utilizing the interlayer coupling in twisted graphene multilayers. Graphene multilayers with interlayer twist occur naturally, when grown   on certain substrates \cite{berger:jpc04,yan:prl12}, or can be manufactured by  stacking graphene single-layers \cite{sanchez:prl12}. For graphene bilayers with a commensurate  interlayer twist Mele has shown \cite{mele:prb10} that the interlayer interaction can open sizable spectral gaps when the sublattice exchange (SE) symmetry is preserved. Here we show that the resulting phase is a mirror TCI, a topological phase distinct from the QSH insulator,  requiring neither spin-orbit interactions nor time-reversal symmetry. To the best of our knowledge it is the first two dimensional TCI in a material that is not superconducting (or superfluid). The gap has been predicted to reach the order of ten ${\rm meV}$ \cite{shallcross:prl08,mele:prb10}. It is thus large enough for this phase to be explored in experiments.  Moreover, 
  gapless boundary excitations, the hallmark of topological phases, can be observed in any crystallographic direction. This is in contrast to generic   TCIs that have gapless modes only in certain high-symmetry directions or planes.  

 { \em Model }:
 To start with we focus on a graphene bilayer with even SE symmetry, as depicted in Fig.\ $ ${1}. 
The low-energy theory of such a bilayer, expanded around the K-points in the Brillouin zone, was derived by Mele in Ref.\ \cite{mele:prb10} and it can be written as
\beq \label{H}
H=v \left(p_S \sigma_x \tau_z + p_{\bar{S}} \sigma_y\right)+ \gamma e^{i \theta l_z\sigma_z \tau_z} l_x e^{-i \theta l_z\sigma_z \tau_z} 
\eeq
\footnote{At high energies there are additional interlayer coupling terms that do not conserve the ``Dirac'' momentum   $\boldsymbol{ p} $ that appears in Eq.\ (\ref{H}) \cite{lopes:prl07}. At the (large) interlayer rotation angles of interest here, however, these terms merely renormalize the  carrier velocity $v$ \cite{lopes:prl07} in the low-energy theory   Eq.\ (\ref{H}) and they thus do not appear explicitly.}.  Here, $\boldsymbol{ p} =  \boldsymbol{\nabla}/i$ (we set $\hbar=1$) is the electron momentum with components  $p_S$  along the mirror axis $S$ in   Fig.\ $ ${1} and  $p_{\bar{S}}$   in the direction perpendicular to it. Furthermore, $v$ is the charge carrier velocity in graphene, and $\gamma$ and $\theta$ parametrize the low-energy interlayer coupling due to lattice commensuration \cite{mele:prb10}. 
 The Pauli matrices $\sigma_i$ act on the  pseudospin of graphene, which distinguishes the two sublattices A and B \cite{neto:rmp09}:  electrons with pseudospin-up reside on sublattice A and  those with pseudospin-down correspondingly  on sublattice B. The Pauli matrices $\tau_i$ act on  the ``valley spin,''  which is up  in band structure valley K and down  in valley K'. Also the   $l_i$ are Pauli marices  and they act on a ``layer spin,'' with spin-up on the top layer and layer spin-down on the lower one. We disregard the electron spin for now.

 \begin{figure}[h]\includegraphics[width=8.5cm] {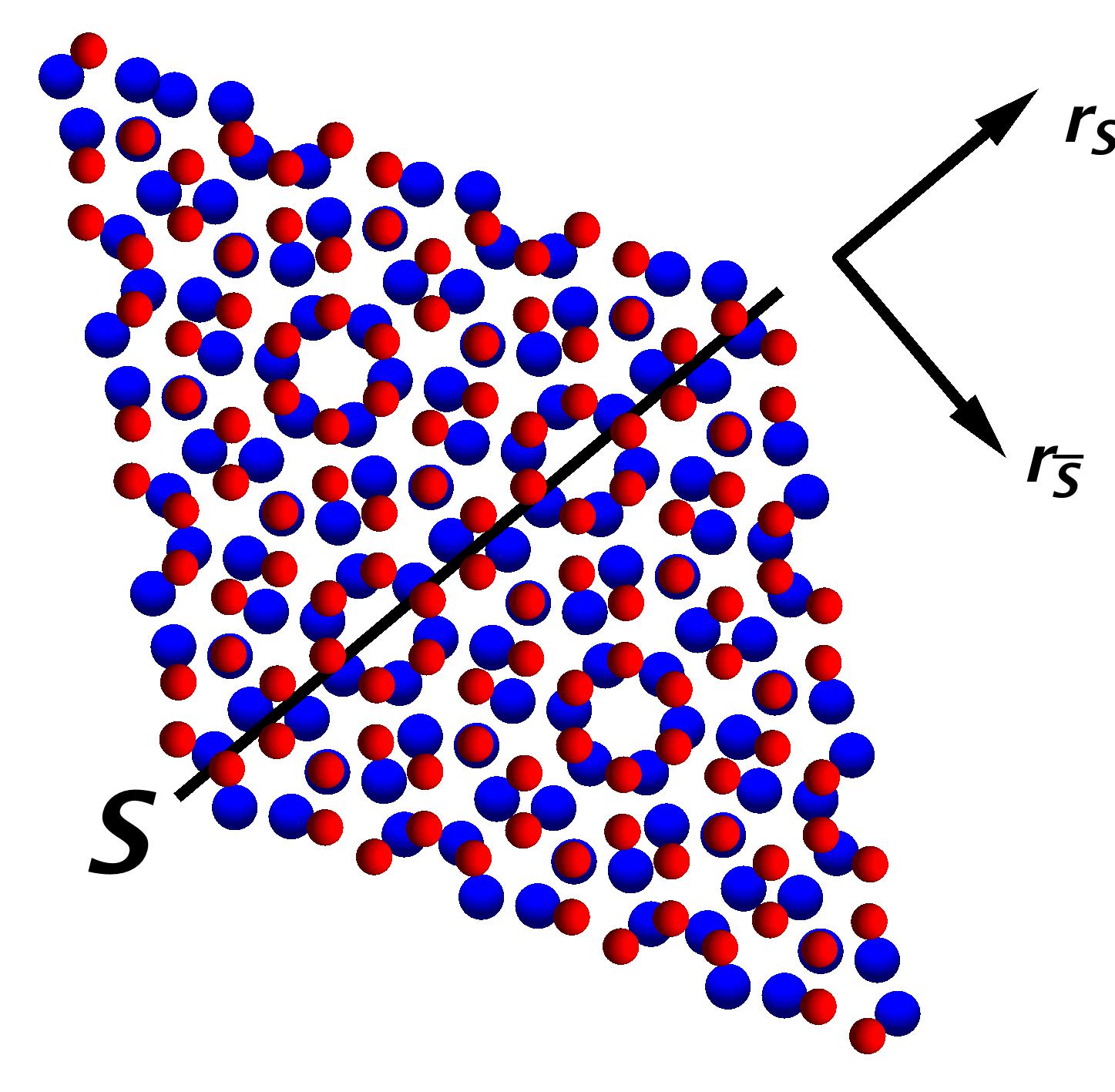}
\caption{SE-even graphene bilayer with an interlayer twist angle of $38.213^{\circ}$. The structure has $C_2$-symmetry on S. For clarity the top layer atoms are depicted smaller than those in the lower layer. }\label{fig1}
\end{figure}

{ \em Symmeries and Topological Classification}:
Insulators can be classified by invariants that characterize  the momentum space topology of   their  Hamiltonian. In the case of two dimensional insulators the generic topological invariant is the first Chern number. Due to their time-reversal symmetry  that characteristic vanishes for the graphene multilayers  considered here.  This means that their Hamiltonians may be deformed continuously and without closing their spectral gap into the Hamiltonian of a simple band insulator.  In the space of all two-dimensional Hamiltonians they are thus said to be topologically trivial. Still, in a restricted space of symmetric Hamiltonians, this need not be so: even though one may be able to deform a Hamiltonian continuously and without gap closings into a trivial one, this might not be possible without breaking one of its symmetries $\Sigma$. In the space of all Hamiltonians with symmetry $\Sigma$ that Hamiltonian is then still topologically nontrivial and one expects gapless boundary modes  in experiments that preserve $\Sigma$. For example, while having zero Chern number,  a QSH Hamiltonian  is topologically nontrivial in the space of all time-reversally invariant Hamiltonians. As a consequence it has the above-mentioned helical edge states. The interlayer phase discussed here is similarly symmetry-protected, but by a crystal symmetry rather than time-reversal symmetry. 
 
The first ingredient to the symmetry that protects the topological phase proposed  here is ${\cal M}$, a $C_2$ symmetry at the axis $S$ in Fig.\ $ ${1}. From a strictly two-dimensional viewpoint, ${\cal M}$ is a combined mirror and interlayer symmetry: mirror reflection on  $S$ and simultaneous layer interchange leaves  an SE-even graphene bilayer unchanged. In the above notation the symmetry ${\cal M}$  has the form
\beq
{\cal M}=M l_x \sigma_x,
\eeq
where $M$  denotes mirror reflection on  $S$ of the spatial coordinates: $M\left(r_S,r_{\bar{S}}\right)=\left(r_S,-r_{\bar{S}}\right)$, cf.\ Fig.\ $ ${1}. 

The second ingredient is a chiral symmetry 
\beq
\tilde{\Sigma} = l_z \sigma_z,
\eeq
  which is due to the particle-hole symmetry of the low-energy theory of the system.  
One readily checks that one has indeed $\{H,\tilde{\Sigma}\}=0$ for the $H$ of Eq.\ (\ref{H}).   We show below that in the space of Hamiltonians $h$ that have the combined chiral and reflection symmetry  
\beq
\Sigma=\tilde{\Sigma} {\cal M},
\eeq
  that is $\{h,\Sigma\}=0$,  the above Hamiltonian $H$   indeed is topologically nontrivial. Within the classification scheme of Ref.\ \cite{chiu:13}, the SE-even  graphene bilayer  is an   $M \mathbb{Z}$-mirror TCI of class BDI. When time-reversal symmetry is broken by additional, ${\Sigma}$-symmetric terms it is of class  AIII.

 { \em  Topological Invariant}:
 The topological arguments below are formulated for lattice models, building on their periodicity in momentum space.  For the purposes of that discussion we therefore analyze a lattice model that has the same low-energy Hamiltonian Eq.\ (\ref{H}) as an SE-even graphene bilayer, but  that has in addition an exact chiral symmetry $\tilde{\Sigma}$ (not only one at low energies):  a tight-binding model of a commensurately rotated graphene bilayer as typically assumed \cite{reich:prb02}, but  with interlayer   hopping only between equal sublattices  (A-to-A and B-to-B).  As shown below, the SE-even graphene bilayer inherits its topological properties from that model.

 The nontrivial topology of this lattice model is characterized by the integer invariant \cite{essin:prb11,chiu:13}
\beq \label{inv}
N_1=\frac{1}{4\pi i}\int_{\rm sBZ} dp_S \,{\rm tr}\,{\Sigma}H^{-1} \partial_{p_S} H\biggr|_{p_{\bar{S}}=0}.
\eeq
Here,   $\partial_\alpha$ denotes the partial derivative with respect to $\alpha$ and the trace ${\rm tr}$ runs over all sites of the supercell defined by the lattice commensuration. Symmetry under translations by supercell lattice vectors is assumed and $p_S$, $p_{\bar{S}}$ are the Bloch momentum eigenvalues in the specified directions.  The integration in Eq.\ (\ref{inv}) is along the line $p_{\bar{S}}=0$   of mirror symmetric momenta. It covers one period in the Brillouin zone of the supercell (sBZ).   An explicit calculation shows that in our model $N_1 =2 \,{\rm sgn}(\gamma \sin 2\theta)$, where ${\rm sgn}\,x$ is the sign of $x$ (cf.\  Appendix A).  The integer nature of  the above invariant is born out by an extension of our model to SE-even graphene multi-layers. For a $2N$-layer as defined in  Appendix A and with an appropriately generalized chiral symmetry  one then finds $N_1 =2N {\rm sgn}(\gamma \sin 2\theta)$.

{ \em  Index Theorem and Boundary Modes}:
The fundamental signature of  topological phases are ungapped modes at boundaries between topologically distinct regions.  For the mirror TCI  of class BDI in two dimensions there is an index theorem \cite{Weinberg:prd81,volovik:03,essin:prb11} that relates $n_{{\Sigma}}$, the sum over the ${\Sigma}$-eigenvalues of all boundary modes at  momentum $p_{\bar{S}}=0$, to the difference between the topological invariants $ N_{1}^{(l)}$ and $ N_{1}^{(r)}$ to the left and to the right of a boundary, respectively:
\beq \label{index}
n_{{\Sigma}}= N_{1}^{(r)}- N_{1}^{(l)}.
\eeq
 It implies a lower bound $| N_{1}^{(r)}- N_{1}^{(l)}|$ on the number of boundary modes between regions with differing topological invariants. Pairs of boundary modes with opposite ${\Sigma}$ eigenvalues evidently do not contribute to $n_{{\Sigma}}$ and can increase the number of boundary modes above   $| N_{1}^{(r)}- N_{1}^{(l)}|$.

 \begin{figure}[h]\includegraphics[width=8.5cm] {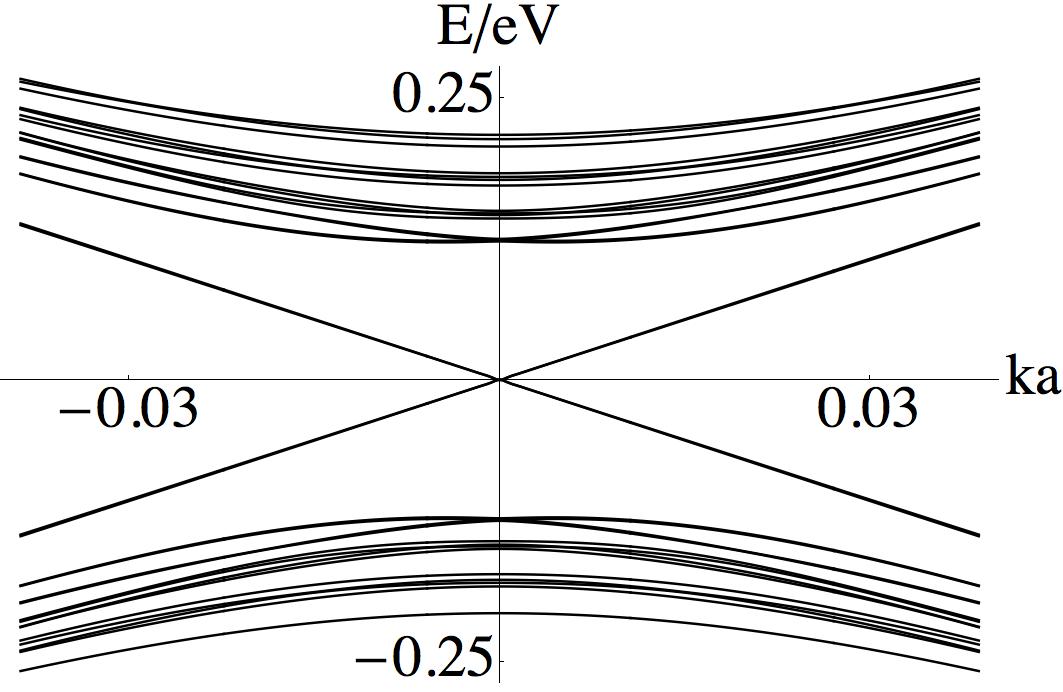} \label{fig2}
\caption{ Spectrum of a 60 unit cells wide ribbon implementing the lattice model described in the main text at an interlayer rotation angle of $38.213^{\circ}$. Here, the edges are perpendicular to the axis  $S$ in Fig.\ 1, so as to preserve  ${\cal M}$-symmetry. The interlayer coupling is $\gamma =300 \,{\rm meV}$. To limit the computational cost we chose a discontinuous dependence of the interlayer coupling on the distance between atoms, in order to enhance the gap due to lattice commensuration \cite{mele:prb10}. This decouples the edge modes of opposing edges already for moderate ribbon widths. Four edge modes are clearly observed, two for each edge of the ribbon. Momentum is measured in units of $1/a$, where $a$ is the lattice constant of graphene.  }
\end{figure}

  In particular at edges, the boundary to the topologically trivial vacuum state with $ N_{1}=0$, such  modes can appear. This occurs when an edge respects the symmetry ${\cal M}$, that is for edges perpendicular to $S$.  According the  index theorem Eq.\ (\ref{index}) there are at least two boundary modes at   such an edge. These gapless modes are clearly observed in the   spectrum of  an ${\cal M}$-symmetric ribbon that implements the above bilayer lattice model, as shown in Fig.\ $ ${2}.

The line $p_{\bar{S}}=0$ cuts through the K-points of graphene and one finds that the topological charge of Eq.\ (\ref{inv}) is concentrated around those points.  Both,  the low-energy theory Eq.\ (\ref{H}) and SE-even bilayer graphene thus indeed inherit their topological properties from the model analyzed above and they share its invariant $N_1$ (see Appendix A).  Because of its rotational invariance   the low-energy  theory Eq. (\ref{H}) is topologically nontrivial not only in the presence of mirror symmetry on $S$, but also for any other mirror axis. 
One therefore expects that topologically protected gapless modes can appear along any crystallographic direction. Such modes may not be observable at an abrupt edge of the system, which may not be described by the low-energy theory Eq.\ (\ref{H}).  By breaking the C$_3$ lattice symmetry abrupt edges may induce  additional interlayer terms. Also, through additional constraints on the low-energy wave functions \cite{akhmerov:prb08} they may break the required mirror symmetry. But gapless boundary modes in arbitrary directions can be observed in systems with smoothly space-dependent parameters, which are entirely described by the Hamiltonian (\ref{H})  with spatially varying couplings. In that case  boundary modes  may be observed for example at lines where the   interlayer coupling $\gamma$ and consequently $N_1$ change sign. In principle, such a sign change would be induced by the  threading of half a magnetic flux quantum through a tube in the gap between the two layers. The index theorem Eq.\ (\ref{index}) predicts at least four boundary modes along such a line. We demonstrate these modes in Fig.\ $ ${3} for a ribbon    with two such flux tubes. To avoid additional edge modes here periodic boundary conditions are applied. In the ribbon whose spectrum is shown in Fig.\ $ ${3} the inserted flux tubes are parallel to S and thus they  break ${\cal M}$-symmetry on the lattice scale. Nevertheless, because of the rotational symmetry of the low-energy theory Eq.\ (\ref{H}), gapless modes are clearly observed.
 In the inset of Fig.\ $ ${3}  we additionally show  the spectrum of the same ribbon  when off-diagonal interlayer hopping (A - to - B and B - to - A) is included, as  in  bilayer graphene. This confirms that SE-even bilayer graphene indeed inherits its topological structure from the model defined above. 

 \begin{figure}[h]\includegraphics[width=8.5cm]{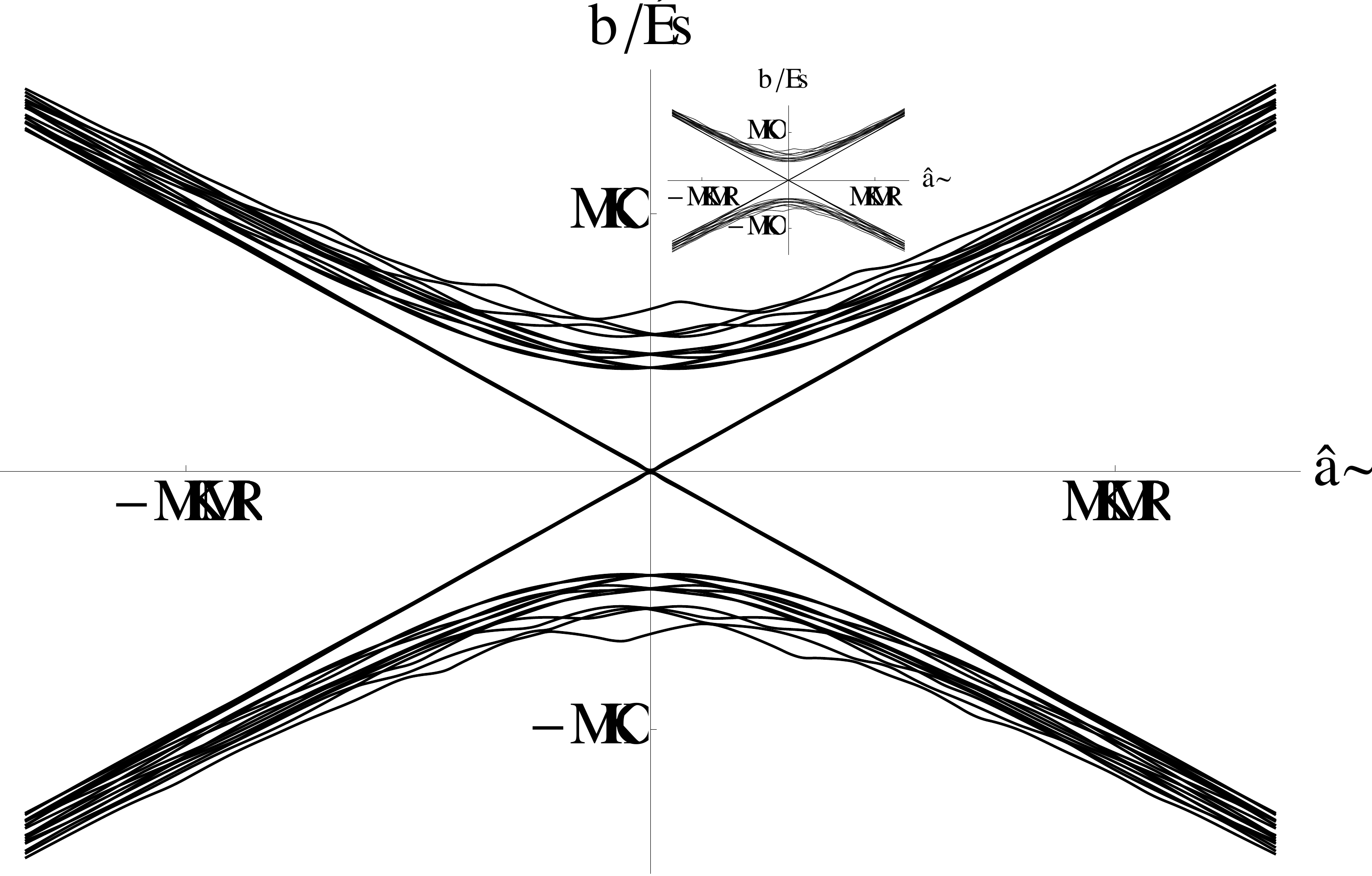}\label{fig3}
\caption{ Spectrum of a  ribbon of the same width as in Fig.\   $ ${2}, but in the crystallographic direction perpendicular to that of Fig.\   $ ${2}. Periodic boundary conditions are applied and two flux tubes parallel to $S$, each of which   induces a sign change in the interlayer coupling $\gamma$, are inserted. The flux tubes break ${\cal M}$-symmetry on the lattice scale, yet  four topological boundary modes per flux tube are  observed. Inset:    The same spectrum for a ribbon with additional pseudospin off-diagonal interlayer hopping (A - to - B and B - to - A), as in bilayer graphene. The graphene bilayer inherits its topological properties from the model described in the text and therefore the same boundary modes are clearly observed. }
\end{figure}

The same boundary modes can also be observed in an experimentally much more accessible set-up:   an SE-even graphene bilayer subject to  an electric potential $V$, which is constant along a line $l$, but confined to within a small distance  $w\ll v/\Delta $ from $l$.  Here, $\Delta= 2 \gamma \sin2\theta $ is the spectral gap of the Hamiltonian (\ref{H}). The potential $V$ has mirror symmetry on   lines $\tilde{S}$ perpendicular to $l$ and due to its rotation symmetry  the Hamiltonian   can be brought into the form of Eq.\ (\ref{H}) with $r_S$ and $r_{\bar{S}}$ replaced by the coordinates $r_{\tilde{S}}$ and $r_{\bar{\tilde{S}}}$ parallel and perpendicular to $\tilde{S}$, respectively. 
We then eliminate    $V$   by the transformation 
\beq
U(r_{\tilde{S}})= e^{-i \phi(r_{\tilde{S}}) \sigma_x \tau_z},
\eeq
where $\phi(r_{\tilde{S}})=\int_{-\infty}^{r_{\tilde{S}}}dr_{\tilde{S}}'\, V(r_{\tilde{S}}')$, at the expense of  a modified interlayer coupling term in the  transformed  Hamiltonian $U^\dag H U$. One finds that if the total phase $\phi(\infty)$  that electrons acquire in the potential $V$ is an integer multiple of $\pi/2$ then far from $l$ the interlayer Hamiltonian after the transformation still has  the form of Eq.\ (\ref{H}). However, if $ \phi(\infty)$  is an odd integer multiple of $\pi/2$  then  in the transformed Hamiltonian the sign of $\theta$ to both sides of $l$ differs.  Consequently  $N_1 $ changes sign across $l$ and at least four boundary modes appear according to Eq.\ (\ref{index}). Close to $l$ there is an extra term    $\Delta \sin(2\phi) {\Sigma}M/2$ in the transformed Hamiltonian, which  breaks the chiral  symmetry ${\Sigma}$. However, while the maximum of that perturbation is of the same order as the gap $\Delta$, at $w\ll v/\Delta$ it is appreciable only in a region  narrow on the scale of the spatial extent of the boundary modes. Therefore the resulting perturbation of the boundary mode energies is small compared to $\Delta$ and in the limit $w\Delta/v\to 0 $ these modes are still gapless. The spectrum displayed in Fig.\ $ ${4} clearly shows these four  boundary modes, again when ${\cal M}$ is broken on the lattice scale by the potential $V$.

 \begin{figure}[h]\includegraphics[width=8.5cm]{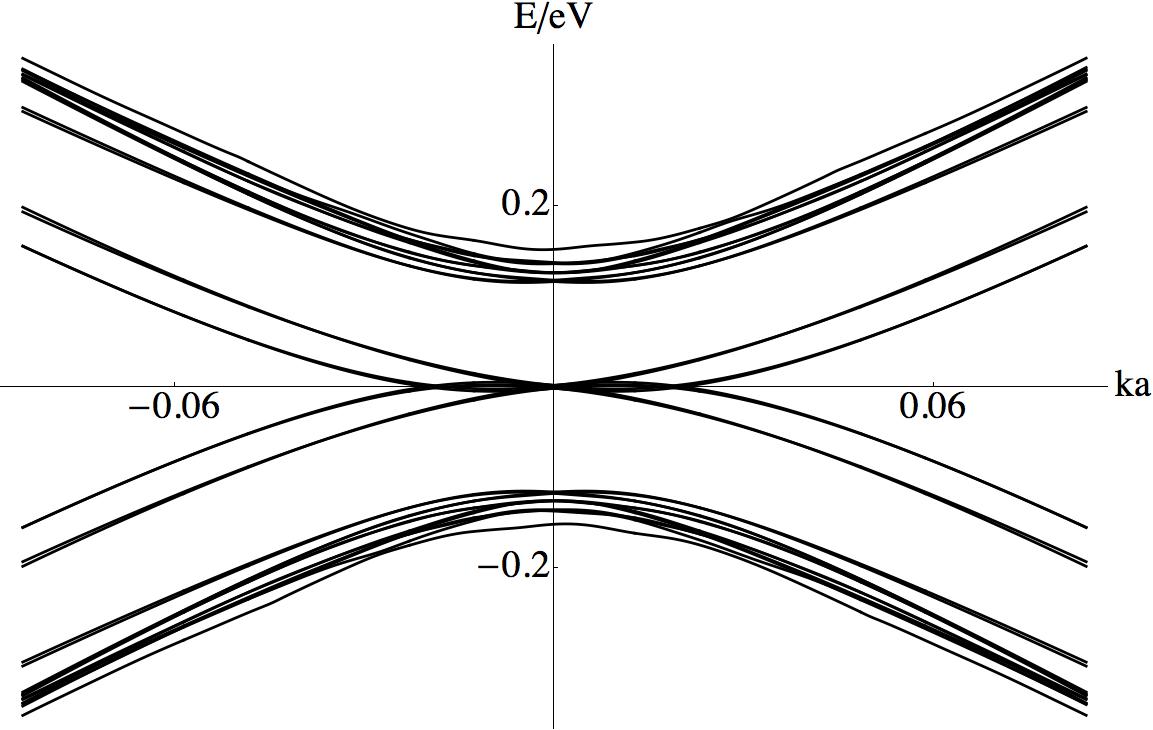}\label{fig4}
\caption{  Spectrum of the same ribbon as shown in Fig.\   $ ${3}, but with the  flux tubes replaced by two line potentials, each of which   induces an effective sign change of the parameter $\theta$ in Eq.\ (\ref{H}). Again, the line potentials break ${\cal M}$-symmetry on the lattice scale, yet  four topological boundary modes per line are  observed.  }
\end{figure}

Finally we briefly comment on the role of the electron spin. Without spin-orbit coupling   the only effect of the electron spin is a doubling of the numbers of boundary modes given above.  Due to the topological stability  of the discussed state this  holds also in the presence of spin-orbit coupling provided that i) that coupling respects $\Sigma$ and that ii)  the corresponding spin-orbit gap is much smaller than the interlayer gap $\Delta$. The intrinsic spin-orbit coupling in graphene, which induces the QSH effect \cite{kane:prl05}, fulfills the first and for typical interlayer twist angles also the second one of these two conditions.  The discussed state is furthermore stable against $\Sigma$-symmetric intervalley scattering and an interlayer bias, as detailed in Appendix B.
  
{\em Conclusion:}  Our discussion identifies a two dimensional TCI in a most widely available material system: graphene multilayers. 
 We have shown that topological boundary modes can be induced electrostatically in arbitrary spatial directions. Besides their fundamental interest these modes implement low-dissipation quantum wires with possible application in graphene electronics. Most importantly, graphene does not pose the experimental challenges of previous implementations of two-dimensional TIs in semiconductor heterostructures. It moreover is accessible to  a number of powerful experimental probes not available for such heterostructures, such as scanning tunneling microscopy and angle resolved photoemission spectroscopy.   It can  therefore be hoped that this proposal  will  significantly advance the development of  two-dimensional TIs with their exciting applications.  

The author thanks P.\ M.\ Goldbart,  P.\ Jarillo-Herrero,  E.\ J.\ Mele, and J.\ Sanchez-Yamagishi  for discussions and gratefully acknowledges  financial support from the NSF  (DMR-1055799).


 \section{ Appendix A}
We first evaluate the invariant $N_1$ in the low-energy theory Eq.\ (\ref{H}). To that end we rewrite  Eq.\ (\ref{inv})  as
\beq \label{inva}
N_1=-\frac{1}{4\pi i}\int_{-\infty}^{\infty} dp_S \,{\rm tr}\,H^{-2}{\Sigma}  \partial_{p_S} H   \times H \biggr|_{p_{\bar{S}}=0},
\eeq
where the trace ${\rm tr}$ now is over the pseudo spin, layer, and valley indices. We could extend the integration to the entire real axis because the integral converges, as is explicitly evident in  Eq.\ (\ref{invb}). 
We next analyze $H$, the last factor in Eq.\ (\ref{inva}), term by term. We note that  all but one term in $H$, Eq.\ (\ref{H}),  anticommute with $H ^{-2}\Sigma  \partial_{p_S}H $ and by the cyclicity of the trace those terms make no contribution to  $N_1 $. The only term in $H $ that does contribute is proportional to  $(\Sigma \partial_{p_S}H )^{-1}$ and one readily finds
\bea \label{invb}
N_1&=&\frac{\gamma\sin2\theta}{4\pi i}\int_{-\infty}^{\infty} dp_S \,{\rm tr}\,H^{-2}  \\
&=&\frac{\gamma\sin2\theta}{4\pi i}{\rm tr}\int_{-\infty}^{\infty}   \frac{dp_S}{p_S^2+2p_S \gamma \tau_z {\cal M}\cos2\theta+\gamma^2 {\cal M}^2} ,\nonumber
\eea
which after diagonalization of ${\cal M}$ evaluates to
\beq \label{N1}
N_1 =2 {\rm sgn}(\gamma \sin 2\theta).
\eeq
As is seen from the convergence of the integral in Eq.\ (\ref{invb}, the topological charge of the low-energy model Eq.\ (\ref{H}) is concentrated around the K-points. One thus expects the same to be true of the lattice model defined in the  main text and consequently that lattice model to share  the invariant   $N_1$ of  Eq.\ (\ref{N1}).  We have confirmed this expectation to hold true  by numerical integration of Eq. (\ref{inv}) over the sBZ for that lattice model.  The low-energy theory Eq.\ (\ref{H}) and therefore also SE-even graphene bilayers thus indeed inherit the topological properties of the lattice model defined in the main text. 
 
There is a straightforward extension of our model to SE-even graphene multi-layers, where $2N$ layers ($N$ is integer) are stacked with alternating, commensurate interlayer rotation angle and coupled by identical hopping amplitudes between nearest layers (the model is readily further generalized to the case of arbitrary nearest-layer couplings and the conclusions are the same as long as a gap opens). Formally, such multi-layers are described by  the same Hamiltonian Eq.\ (\ref{H}), but with different definitions of the layer-spin operators: in the $2N$-layer spin space they have matrix elements $\left(l_x\right)_{j,k}=\delta_{|j-k| ,1}$ and $\left(l_z\right)_{j,k}=(-1)^{j}\delta_{j,k}$. The Hamiltonian then still respects the chiral  symmetry ${\Sigma}$ (modified accordingly)  and its invariant $N_1 $ is calculated in the same way as for the bilayer. The only difference is that the involved matrices  are now higher dimensional, which increases the trace, such that $N_1 =2N {\rm sgn}(\gamma \sin 2\theta)$. This exemplifies that the invariant $N_1$   indeed takes values in  $\mathbb{Z}$.

\section{ Appendix B}
 
The discussed topological state is distinct from the marginal topological phase in Bernal stacked graphene bilayers \cite{li:prb10}, which was discussed first in Ref.\ \cite{martin:prl08} and later termed  ``quantum valley Hall insulator.'' The latter phase is topologically nontrivial only at low energies and after singling out one of the two band structure valleys.  In that phase the winding numbers in the  two valleys of graphene cancel, while in the state considered here they add up.   In contrast to the quantum valley Hall insulator the  interlayer phase discussed here therefore is robust against intervalley scattering. To illustrate this we add to  the Hamiltonian Eq. (\ref{H}) a term that scatters between valleys. This is done  through an ${\cal M}$-symmetric  Kekule distortion of the hopping amplitudes in our bilayer lattice model, with relative amplitude $u_{\rm Kekule}$. As is born out in the phase diagram of Fig.\ $ ${5}, the above interlayer phase with  $|N_1 |=2$ indeed extends far into the region with nonzero $u_{Kekule}$. Furthermore, we note that the predicted phase does not require that the chiral  symmetry $\tilde{\Sigma}$ and the reflection/interlayer symmetry ${\cal M}$ are satisfied separately. Only their combination ${\Sigma}$ needs to be respected. In particular the discussed phase is robust against an    interlayer bias $V_b$, a very common perturbation of graphene bilayers, which is described by a term $ V_b\l_z$ in the Hamiltonian. Nonzero $V_b$ breaks both,  ${\cal M}$  and  $\tilde{\Sigma}$ separately, but it respects the combined  symmetry ${\Sigma}$. Accordingly  the above mirror topological phase  persists also at nonzero $V_b$. In fact, an interlayer bias tends to stabilize that phase, as seen in the phase diagram of Fig.\ $ ${5}.

  \begin{figure}[h]\includegraphics[width=8.5cm]{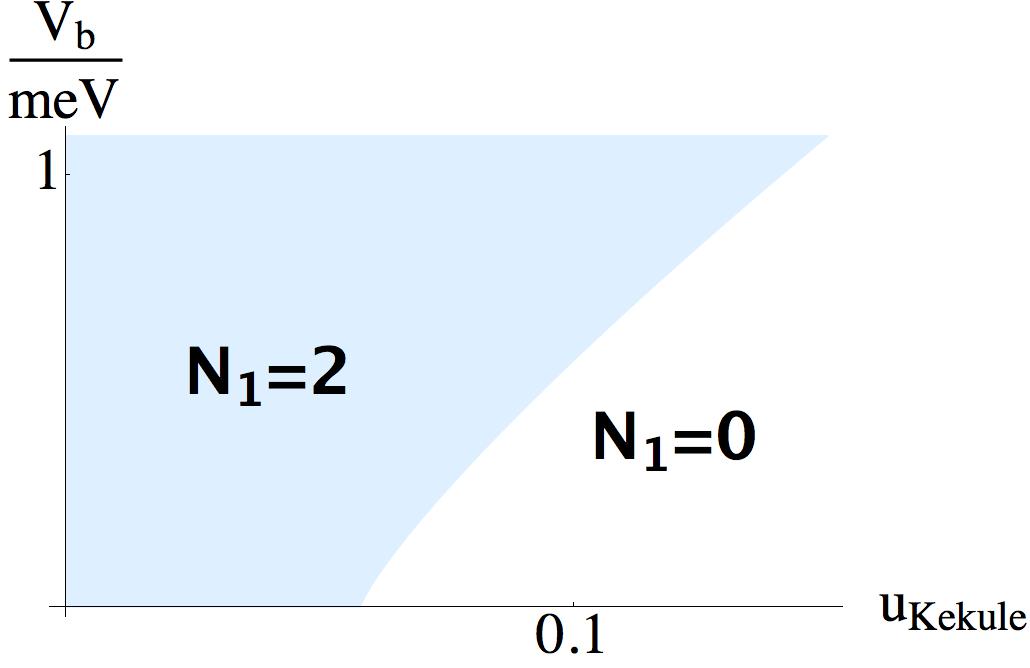}\label{fig5}
\caption{ Phase diagram of an SE-even bilayer with  interlayer bias $V_b$ and ${\cal M}$-symmetric Kekule-distortion of the hopping amplitudes, which has relative amplitude $u_{\rm Kekule}$. The discussed interlayer phase $N_1=2$  persists into the region with intervalley scattering and interlayer bias.  }
\end{figure}

\end{document}